\documentclass[a4paper,12pt]{spieman}  
\usepackage{amsmath,amsfonts,amssymb}
\usepackage{graphicx}
\usepackage{setspace}
\usepackage{tocloft}
\usepackage{lineno}
\title{Room temperature valley coherence in monolayer WSe$_2$ mediated by chiral nematic liquid crystal}

\author[a,b]{Hassan Lamsaadi}
\author[a]{Andrea Balocchi}
\author[b]{Aurélien Cuche}
\author[b]{Hugo Lourenco Martins}
\author[b]{Vincent Paillard}
\author[b]{Sébastien Weber}
\author[b,*]{Jean Marie Poumirol}
\author[b,*]{Gonzague Agez}
\affil[a]{Université de Toulouse, CNRS, CEMES, Toulouse, France}
\affil[b]{LPCNO, INSA, Toulouse, France}

\cftpagenumbersoff{figure}
\cftpagenumbersoff{table} 
\begin{document}
\maketitle

\begin{abstract}
Valley coherence refers to a phase‑coherent superposition of inequivalent momentum valleys, in which quantum information can be encoded in the relative valley phase. Chiral nematic liquid crystals, by imposing a flip of the spin angular momentum upon light reflection, provide an effective photonic environment for optically coupling excitons of the K and K' valleys in monolayer semiconducting transition‑metal dichalcogenides. We experimentally demonstrate that using such liquid crystal as a  substrate, it is possible through nearfield interaction to engineer a room temperature mechanism for inducing the intervalley coupling. Our results show that this approach provides a simple and scalable route toward valleytronic functionalities based on controlled coherent emission from valleys with opposite Berry curvature.
\end{abstract}


{\noindent \footnotesize\textbf{*}Jean Marie Poumirol,  \linkable{jean-marie.poumirol@cemes.fr} }
{\noindent \footnotesize\textbf{*}Gonzague Agez,  \linkable{gonzague.agez@cemes.fr} }

\begin{spacing}{2}   

\section{Introduction}

Inter‑valley coherence in monolayer semiconducting transition‑metal dichalcogenides (1L-TMD) refers to the formation of a well‑defined quantum superposition between excitonic states residing in the K and K' valleys of the Brillouin zone\cite{jones2013optical,hao2016direct,gupta2023observation,cadiz2022imaging}. Such coherent superpositions form the basis of valley‑dependent optical phenomena and constitute a promising resource for information processing in atomically thin semiconductors\cite{unuchek2018room,schaibley2016valleytronics,rivera2018interlayer}. However, maintaining a stable phase relation between valleys at room temperature remains intrinsically challenging. Long‑range electron–hole exchange induces rapid pseudospin precession, while disorder, phonon scattering, and momentum‑dependent dephasing further suppress coherence on sub‑picosecond timescales~\cite{mak2016photonics, xu2014spin}. As a result, in 1L-TMD robust intervalley coherence has so far been observed predominantly under cryogenic conditions~\cite{zhu2014anomalously} and/or through highly resonant excitation schemes~\cite{mak2012control, zeng2012valley, cao2012valley}. A key question is whether these valley pseudospins can be coherently controlled, in the same way as real spins, to enable future quantum technologies.

In response, several photonic strategies have been developed to overcome these limitations by engineering the light–matter interaction experienced by valley excitons. Early approaches exploited the intrinsic spin–valley locking of excitons, but provided only limited control over the pseudospin dynamics~\cite{smolenski_tuning_2016}. More advanced platforms then emerged, relying on strongly structured photonic environments, including anisotropic microcavities, chiral metasurfaces with nanoscale patterning~\cite{guddala_valley_2019, loren_circular_2023}, topological photonic lattices supporting spin–momentum‑locked modes~\cite{jimenez-galan_lightwave_2020},or graphene-TMD heterostructures \cite{lorchat2018room }. Within this broader class of engineered photonic architectures, breaking the vertical symmetry of the cavity has also been shown to induce an effective in‑plane spin–orbit field, described as a photonic Rashba‑like effect, which couples selectively to K and K' excitons~\cite{rong_spin-valley_2023, rong_photonic_2020}. 

Despite the significant progress enabled by these photonic architectures, valley coherence is typically achieved through carefully engineered photonic band structures, symmetry breaking, or spatial confinement. As a result, device performance remains highly sensitive to fabrication imperfections and difficult to sustain under realistic, room‑temperature conditions. This accumulation of constraints underscores a broader challenge: achieving robust intervalley coherence at room temperature while maintaining a platform that is simple, tolerant to fabrication variations, and compatible with large‑scale integration.

It is precisely in this context that chiral nematic liquid crystals (CLC) emerge as a particularly powerful alternative, both conceptually and experimentally. Due to their intrinsic helical molecular order, CLCs exhibit a circularly polarized Bragg reflection that selectively reflects one helicity while reversing the spin angular momentum of the reflected light. This built‑in spin‑flip process provides a natural photonic spin–orbit interaction that mediates optical coupling between the K and K' valleys, without the need for elaborate nanofabrication or finely engineered device geometries.

In this study, we focus on the photoluminescence response of a WSe$_2$ 1L-TMD placed directly on a left‑handed CLC substrate in a planar configuration. This system provides an ideal platform for probing how chiral Bragg reflection can reshape light–matter interactions in 2D semiconductors. To rigorously identify the role of the chiral photonic environment, we compare three well‑defined configurations: a WSe$_2$ monolayer on a CLC whose polarization selective photonic bandgap (PBG) spectrally overlaps the excitonic emission of WSe$_2$, a WSe$_2$ monolayer on a CLC whose BPG lies outside the excitonic emission range, and a reference monolayer deposited on an inert glass substrate. This set of controlled conditions enables us to isolate the specific contribution of the chiral Bragg mechanism to the emission properties of WSe$_2$.

\section{Results}
\begin{figure*}[ht!]
    \centering
    \includegraphics[width=\textwidth]{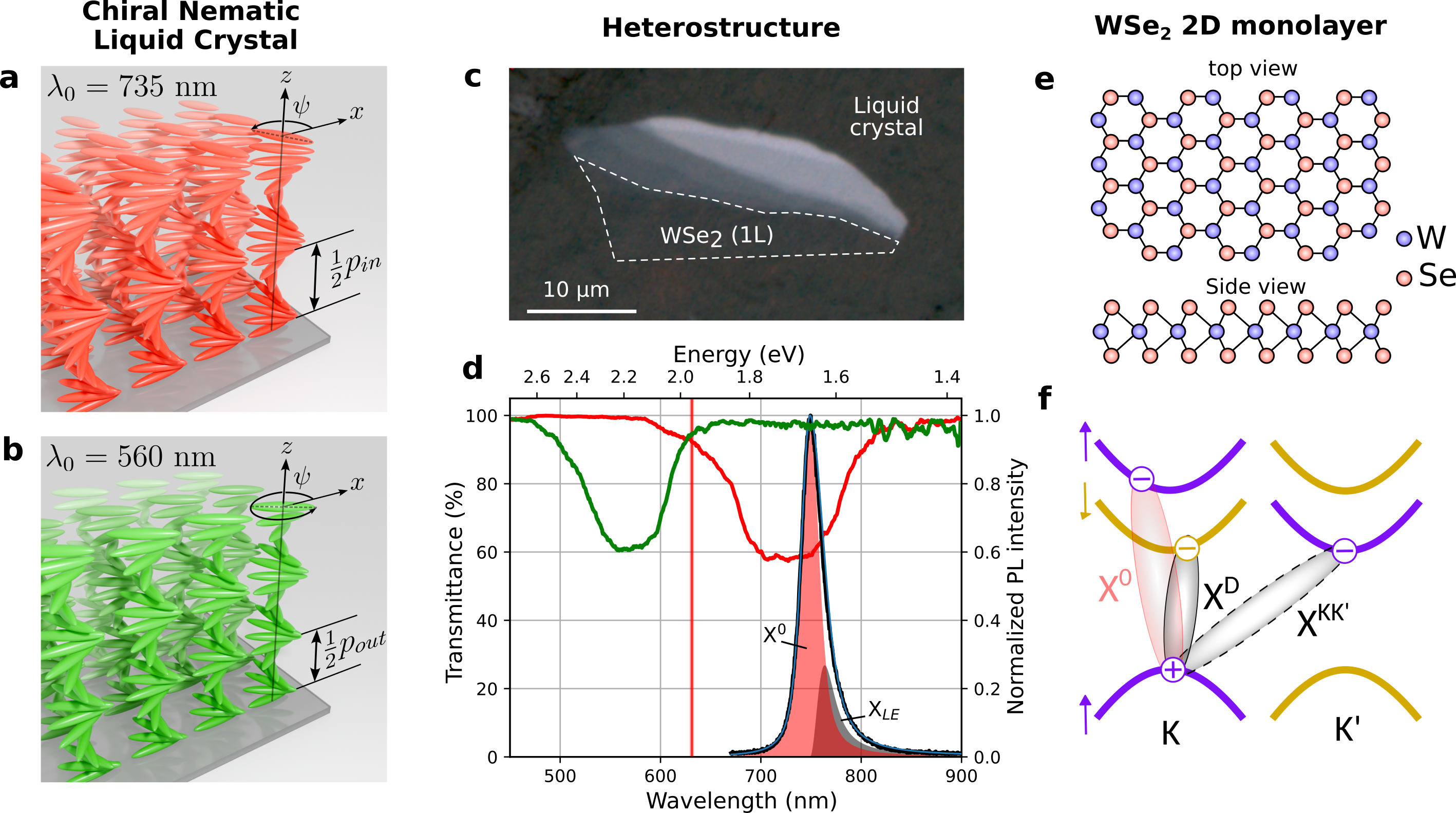} 
    \caption{\textbf{Sample preparation and characterization.} 
    (\textbf{a}-\textbf{b}) Schematic of 3D structure of the CLCs with PBG centered at 735nm and 560nm, respectively Sample \textsc{in} and \textsc{out}. 
    (\textbf{c}) Optical micrograph of WSe$_2$-CLC heterostructure. The monolayer region is indicated by the dashed white lines. 
    (\textbf{d}) Typical PL spectrum of WS$_2$ monolayer showing dominant bright exciton contribution and a lower energy sideband associated mainly to spin-dark excitons. (\textbf{e}) Top and side views of a TMD monolayer chemical struct. 
    (\textbf{f}) Electronic picture of bright (X$^0$) A-exciton and both spin- and momentum forbidden dark excitonic states in WSe$_2$ monolayer. }
    \label{fig:Fig.1}
\end{figure*}

Samples were prepared by depositing a CLC thin film onto a glass substrate in a planar (Grandjean) configuration. A monolayer (ML) of WSe$_2$, mechanically exfoliated from a commercial bulk crystal, was subsequently transferred onto direct contact with the free up interface of the CLC (see Methods). An optical micrograph of a representative device is shown in Fig.~\ref{fig:Fig.1}\textbf{c}, where the ML region is indicated by dashed outlines.

The central wavelength $\lambda_0$ of the CLC PBG is related to the helical pitch $p$ according to
\begin{equation}
\lambda_0 = \bar{n} \, p \cos \theta ,
\label{eq:bragg_condition}
\end{equation}
where $\theta$ is the angle between the incident light and the helical axis. The average refractive index is defined as
$\bar{n} = (n_e + n_o)/2$, with $n_e$ and $n_o$ denoting the extraordinary and ordinary refractive indices of the liquid crystal, respectively. The helicoïdal structure is characterized by the local twist angle
\begin{equation}
\psi(z) = \frac{2\pi \chi z}{p} ,
\end{equation}
which quantifies the torsion along the $z$ axis normal to the glass substrate. Here, $\chi = \pm 1$ specifies the handedness of the supramolecular helix. The pitch $p$ is tuned during fabrication by controlling the relative concentration of the chiral component.

To disentangle the role of the circularly polarized PBG from effects solely arising from the presence of the CLC beneath the  TMD layer, exfoliated monolayer WSe$_2$ have been transferred on two distinct CLC substrates (see method). The first sample, hereafter referred to as sample \textsc{in}, exhibits a PBG (defined at half maximum) spanning the 660--780~nm spectral range, chosen to spectrally overlap with the PL emission band of monolayer WSe$_2$. In contrast, the second substrate, denoted Sample \textsc{out}, displays a Bragg reflection band between 520 and 610~nm, resulting in no spectral overlap with the WSe$_2$ PL.

The corresponding PBGs are visible in the normal-incidence transmittance spectra shown in Fig.~\ref{fig:Fig.1}\textbf{d}, recorded using unpolarized incident light. Under these conditions, the maximum depth of the PBG in transmittance is intrinsically limited to 50\% of the incident intensity. This originates from the polarization-selective nature of Bragg reflection in CLC, which reflects only the circularly polarized component whose helicity $\sigma$ matches the handedness of the supramolecular helix, namely $\sigma = -\chi$, while the opposite circular polarization $\sigma = +\chi$ is transmitted. Here, $\sigma = + 1$ ($\sigma = - 1$) denotes the helicity of a left- (right-) handed circular polarization of light.

In both configurations, the excitation wavelength ($\lambda_\mathrm{exc} = 638$~nm) was deliberately chosen to lie outside the photonic band gaps, thereby excluding excitation‑related effects. This comparative approach enables disentangling PBG‑induced modifications of the monolayer optical response from those arising from the CLC environment.
A typical room-temperature photoluminescence (PL) spectrum acquired under pulsed laser excitation at low average power ($P_{laser} \approx 10\mu $W, $\lambda_{\mathrm{laser}} = 638$nm) is presented in Figure \ref{fig:Fig.1}\textbf{d}. The PL emission is dominated by the A-exciton resonance centered at $\approx 1.66$ eV. This feature may arise from the spectral overlap of spin-allowed and spin-forbidden A-excitons, which are separated in energy by about $40\,meV$\cite{zhou2017probing,lamsaadi2025exciton} (see Fig. \ref{fig:Fig.1}\textbf{f}). As a result, the PL spectrum is asymmetric, exhibiting a low-energy sideband $X_\text{LE}$ primarily associated with contributions from spin-dark excitons. Others excitonic species may contribute to the observed sideband, including intervalley excitons (e.g., $X^{KK'}$)\cite{brem2020phonon} and trion emission originating from residual charge carriers, expected $\approx 30-40 meV$\cite{li2018revealing} below the spin-allowed A-exciton energy. In the following, we focus exclusively on the bright 
A-exciton contribution, which is extracted from the total PL spectrum via a Lorentzian lineshape fit, allowing us to deconvolve it from the 
low-energy sideband (see Fig. \ref{fig:Fig.1}\textbf{d}).

\begin{figure*}[ht]
    \centering
    \includegraphics[width=\textwidth]{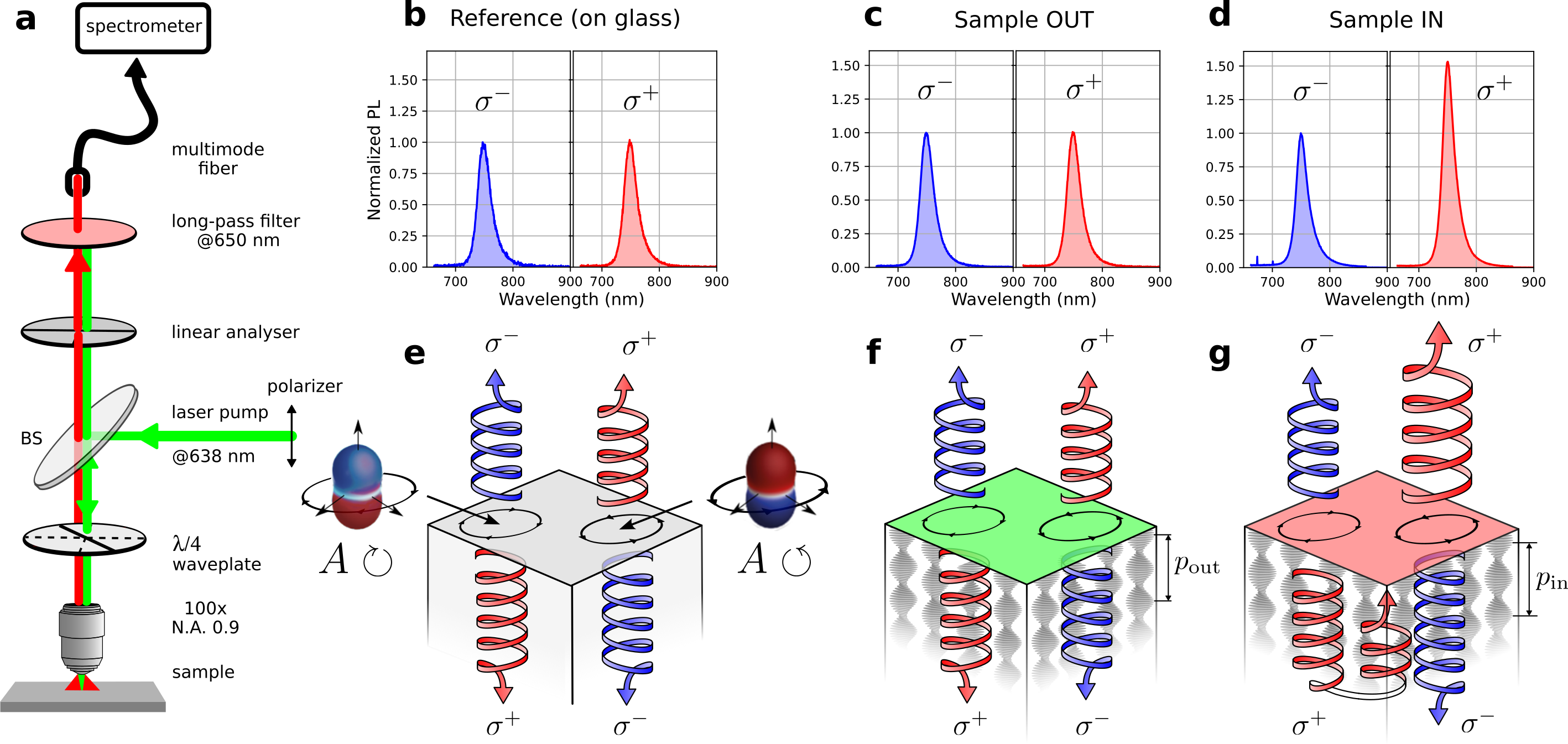} 
    \caption {\textbf{Circularly polarized photoluminescence.}
(\textbf{a}) Schematic of the experimental setup used for circularly polarized PL measurements. The excitation beam is focused onto the sample using a $\times100$ objective with a numerical aperture of 0.9.
(\textbf{b}--\textbf{d}) Circularly polarized PL spectra of WSe$_2$ monolayer for three different configurations. Blue (red) curves correspond to right- (left-) handed circularly polarized emission.
(\textbf{b}) Reference sample : monolayer deposited on a glass substrate.
(\textbf{c}) Sample \textsc{out}: monolayer deposited on a left-handed CLC with pitch $p_{\mathrm{out}}$.
(\textbf{d}) Sample \textsc{in}: monolayer deposited on a left-handed CLC with pitch $p_{\mathrm{in}}$.
(\textbf{e}--\textbf{g}) Schematic illustrations corresponding to the configurations in (\textbf{b}--\textbf{d}), respectively. The emission of $A\circlearrowright$ and $A\circlearrowleft$ excitons, associated with the K and K' valleys, is depicted. The radiation patterns of the excitons in free space are shown in (\textbf{e}). In (\textbf{f}) and (\textbf{g}), the chiral nematic structure is represented with the distinct pitches $p_{\mathrm{out}}$ and $p_{\mathrm{in}}$.}
    \label{fig:Fig.2}
\end{figure*}

\subsection{Circularly polarized photoluminescence}

To characterize the influence of the CLC substrate on the polarization state of the light emitted by exciton recombination, we performed circular-polarization-resolved photoluminescence measurements. As described in Fig.\ref{fig:Fig.2}\textbf{a}, the excitation laser is circularly polarized (left or right), and the emitted photoluminescence is analyzed on a left/right orthogonal basis using a combination of a quarter-wave plate and a linear polarizer. The circularly polarized emission spectra shown in Fig.~\ref{fig:Fig.2}\textbf{b}–\textbf{d} are obtained by summing two spectra recorded for orthogonal orientations of the $\lambda/4 $ wave plate. This procedure equally averages right- and left-handed circular non-resonant excitation, thereby suppressing contributions purely related to the excitation conditions. The resulting detection scheme implements the calibration-free protocol described in the Supplementary Information, providing direct and robust access to the circular polarization populations $I_{\sigma^+}$ and $I_{\sigma^-}$. The averaging procedure serves solely to correct for instrumental collection imbalances between the two circular polarization channels, and does not affect the physical content of the PL signal. Indeed, since all measurements are performed under off-resonance excitation, the polarization memory of the excitation laser is largely lost upon carrier relaxation, and the detected circularly polarized PL signal is therefore independent of the excitation helicity. One should notice that the PL light is collected up-ward in free space, \textit{i.e.} not transmitted through the CLC.

As a reference, we first measured the PL properties of WSe$_2$ monolayer deposited on glass. As expected for such a configuration at room temperature\cite{yan2015valley,huang2016probing}, the $\sigma = \pm 1$ contributions to the PL response are identical. 
The PL measured on sample \textsc{out} is displayed in figure \ref{fig:Fig.2}\textbf{c} and shows no observable circular dichroism. This observation indicates that the liquid‑crystal environment, by itself, does not lift the natural balance between right‑ and left‑handed circular emission. In contrast, for sample \textsc{in} (Figure \ref{fig:Fig.2}\textbf{d}), when the CLC pitch is adjusted to match the energy of the emitted PL, a clear difference is observed between the $\sigma = \pm 1$ spectra, with a degree of circular polarization $DoPC$ reaching values of 0.2, with $DoCP = (I_{\sigma^+} - I_{\sigma^-}) / (I_{\sigma^+} + I_{\sigma^-})$.

Bright excitons in 1L-TMDs are spinning dipoles, i.e., they can be described as a superposition of two perpendicular electric dipoles with a relative phase shift of $\pm$ $\pi$/2, thus they emit circularly polarized light. TMDs exhibit chiral optical selection rules, meaning that right- and left-rotating excitons (A$\circlearrowright$, A$\circlearrowleft$) populate two distinct valleys (K and K’ respectively). As illustrated in Fig.~\ref{fig:Fig.2}\textbf{e–g}, for a given excitonic recombination, the emission is $\sigma=+1$ in the upward direction and $\sigma=-1$ in the downward direction; for the exciton in the opposite valley, this helicity pattern is reversed.

As illustrated in Figure \ref{fig:Fig.2}\textbf{f}, out of the Bragg gap, both $\sigma = \pm 1$ downward emited light is able to pass through the substrate as it is the case for more conventional substrate such as glass (Fig. \ref{fig:Fig.2}\textbf{e}). In absence of external coupling process, no circular dichroism is observed in the PL of the reference sample, reflecting the energetic degeneracy of the K and K' valleys and the rapid intervalley scattering that suppresses any steady‑state valley polarization at room temperature.\cite{yan2015valley,huang2016probing}.

In contrast, when light is emitted inside the PBG, a circular Bragg reflection occurs, preserving the electric field helix handedness as schematically represented in Figure \ref{fig:Fig.2}\textbf{g}. Hence the spin angular momentum projection along the $z$ axis per photon is flipped from $s_z = -\chi \hbar$ to $s_z = +\chi \hbar$ \cite{rafayelyan2017ultrabroadband, barboza2016berry}. The $\sigma = 1$ light is no longer transmitted but is reflected conserving his $\sigma =1$ nature upon reflection. When an equal amount of A$\circlearrowleft$ and A$\circlearrowright$ are recombining by unit of time, the resulting total contribution of the left hand light measured in the far optical field upward is therefore enhanced. 

Although this net dichroism does not correspond to a truly chiral light source in the global sense, as the integrated chirality over $4\pi$ vanishes, it results from the breaking of the $\pm z$ symmetry by the CLC \cite{wang2019fully}. Importantly, this approach still enables the generation of chiral light in a chosen direction under ambient conditions.

\begin{figure*}[t!]
    \centering
    \includegraphics[width=\textwidth]{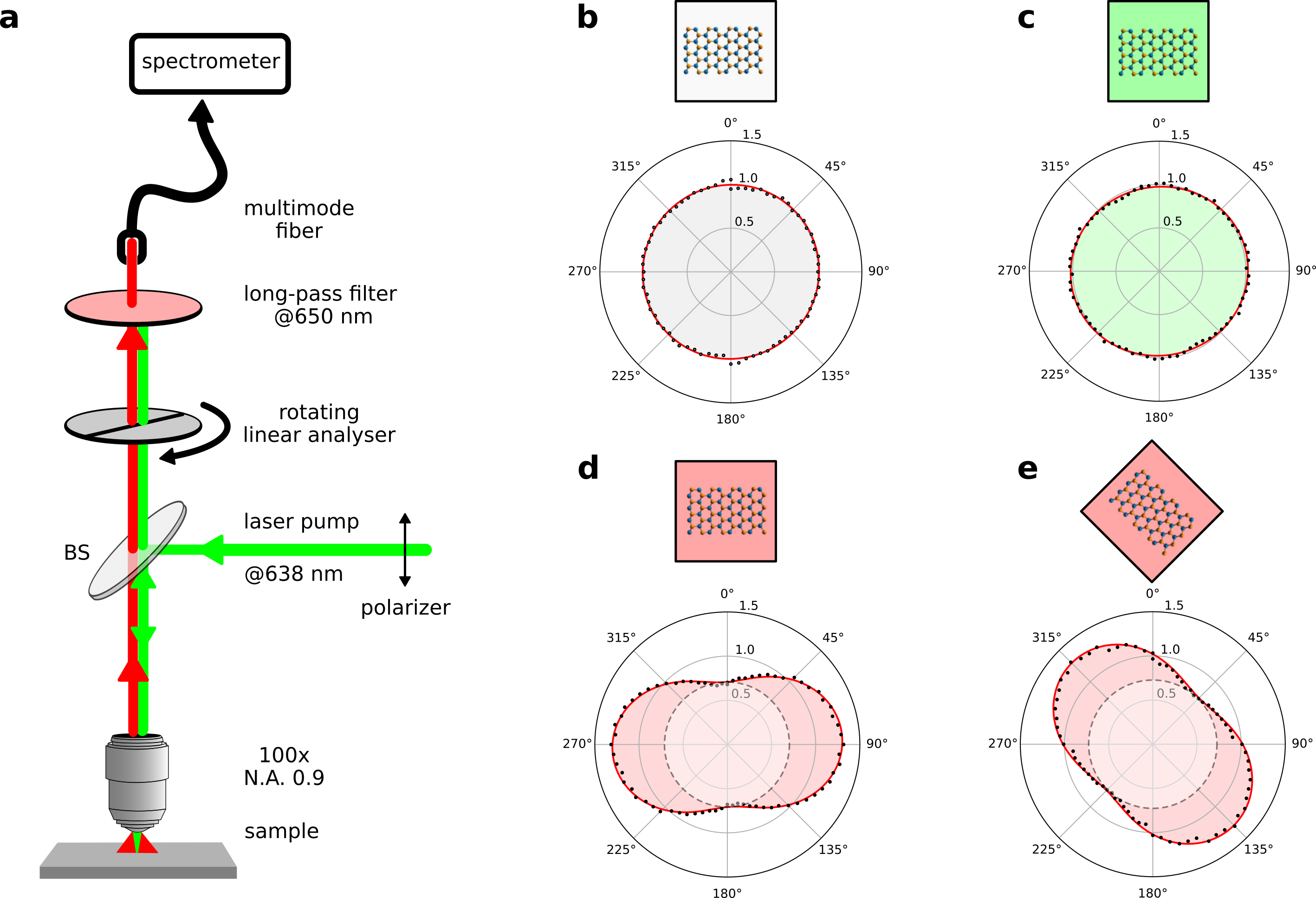} 
    \caption{\textbf{Experimental demonstration of inter-valley coherence.}
(\textbf{a}) Schematic of the polarization-resolved PL setup.
(\textbf{b}--\textbf{e}) Top: schematic of the TMD monolayer deposited on different substrates. Bottom: corresponding polar plots of the emitted PL polarization state.
(\textbf{b}) Reference configuration with the TMD monolayer deposited on a glass substrate.
(\textbf{c}) WSe$_2$ monolayer on CLC in the \textsc{out} configuration
(\textbf{d}) Same system in the \textsc{in} configuration,  leading to the emergence of a pronounced linear polarization signature associated with inter-valley coherence.
(\textbf{e}) Same as (\textbf{d}), but with the substrate rotated by an angle $\alpha$ around the optical axis. In (\textbf{d}) and (\textbf{e}), dashed circles indicate the isotropic reference level.}
    \label{fig:Fig.3}
\end{figure*}

\subsection{Inter-valley coherence}

Strikingly, polarization-resolved photoluminescence measurements reveal a profound modification of the polarization state of the PL. The experimental configuration, schematically shown in Figure \ref{fig:Fig.3}\textbf{a}, relies on a rotating linear polarizer placed in the detection path, allowing us to reconstruct the polarization state of the PL through polar plots of the detected intensity. Four representative polarization diagrams are displayed in Figure \ref{fig:Fig.3}\textbf{b-e}.

For the reference monolayer deposited on the glass substrate (Fig. \ref{fig:Fig.3}\textbf{b}), the polar diagram is perfectly circular, indicating the absence of any privileged linear polarization axis. The emitted light is therefore fully depolarized in the linear basis, consistent with an incoherent superposition of $\sigma = +1$ and  $\sigma = -1$ components. A similar behavior is observed for the \textsc{out} configuration (Fig. \ref{fig:Fig.3}\textbf{c}). The circular polar diagram obtained in this case demonstrates that no linear polarization is induced by the substrate itself, ruling out trivial optical or material anisotropies as the origin of the effects discussed below.

In sharp contrast, when the PL emission is spectrally positioned within the Bragg band, the polar diagram exhibits a two-lobe peanut-shaped pattern (Fig. \ref{fig:Fig.3}\textbf{d}). Quantitative analysis shows that about 30\% of the emitted intensity is linearly polarized, providing direct experimental evidence of intervalley coherence in monolayer WSe$_2$ persisting at room temperature. In this regime, the strong linear polarization component evidences that the $\sigma =+1$ and $\sigma =-1$  PL components are phase-locked. This observation constitutes the central experimental result of this work.

Finally, Figure \ref{fig:Fig.3}\textbf{e} shows the polar diagram obtained in the \textsc{in} configuration after rotating the sample. The linear polarization lobes co-rotate rigidly with the sample, demonstrating that the orientation of the linear polarization component is directly tied to the optical properties of the underlying substrate. This observation unambiguously establishes a coupling between the linearly polarized PL and the strongly optically anisotropic environment.

\subsection{Coupling process}

The optical response of the system is described by the coupled dynamics of the excitonic polarizations in the two inequivalent valleys K and K'. Excitons in the K valley emit circularly polarized photons that, upon Bragg reflection, undergo a spin flip and are selectively reinjected into the K' valley. As a result, the optical coupling between valleys is intrinsically unidirectional, from K to K'.

This situation can be modeled by the following Maxwell-Bloch formalism :

\begin{align}
\frac{\mathrm d E_K}{\mathrm d t}
&=
\left( i \omega_0 - \gamma \right) E_K
+ g P_K
\\[4pt]
\frac{\mathrm d E_{K'}}{\mathrm d t}
&=
\left( i \omega_0 - \gamma \right) E_{K'}
+ g P_{K'}
+ \beta E_K
\\[4pt]
\frac{\mathrm d P_{K}}{\mathrm d t}
&=
\left( i \Delta - \gamma_\phi \right) P_{K}
+ g E_{K}
\\[4pt]
\frac{\mathrm d P_{K'}}{\mathrm d t}
&=
\left( i \Delta - \gamma_\phi \right) P_{K'}
+ g E_{K'}
\end{align}

Where $E_{K}$ and $E_{K'}$ are the slowly varing amplitude of the optical field associated with the valley K and K' respectively, and $P_{K}$ and $P_{K'}$ are the corresponding coherent excitonic polarizations. The frequency $\omega_0$ is fixed by the A‑exciton resonance and not by a cavity mode, as the optical feedback does not form a well‑defined resonator. The parameter $\gamma$ describes the photon loss rate, including radiative decay. The associated lifetime $\tau_L$ is estimated from time-resolved spectroscopy to be on the order of a few hundred picoseconds \cite{cadiz2018exciton,godde2016exciton,yan2014photoluminescence}.
The excitonic polarizations evolve at the exciton–photon detuning $\Delta$ and decay at the dephasing rate $\gamma_\phi$, which accounts for coherence loss of the excitonic dipoles. According to the reference \cite{wang2025direct}, this process is in the range of 0.2-0.3 ps in WSe$_2$ monolayer.
The terms proportional to $g$  describe the coherent exciton–photon coupling within each valley, assumed to be valley conserving (i.e. without inducing intervalley coupling) and identical for K and K'.
The term $\beta E_K$ in the equation for $E_{K'}$ represents a unidirectional optical coupling from the K to the K' valley, arising from the reflection of circularly polarized photons by the Bragg structure accompanied by a spin‑flip process. The absence of a reciprocal term in the equation for $E{_K}$ reflects the intrinsically non‑reciprocal nature of this feedback‑mediated coupling.
Since the field $E_{K'}$ is populated through the coherent transfer $\beta E_K$, the induced polarization $P_{K'}$ can acquire a well‑defined phase only if this phase‑locked injection occurs on a timescale shorter than the excitonic dephasing time, leading to the key necessary condition:
\begin{equation}
\beta > \gamma_\phi
\end{equation}

\begin{figure*}[t!]
    \centering
    \includegraphics[width=\textwidth]{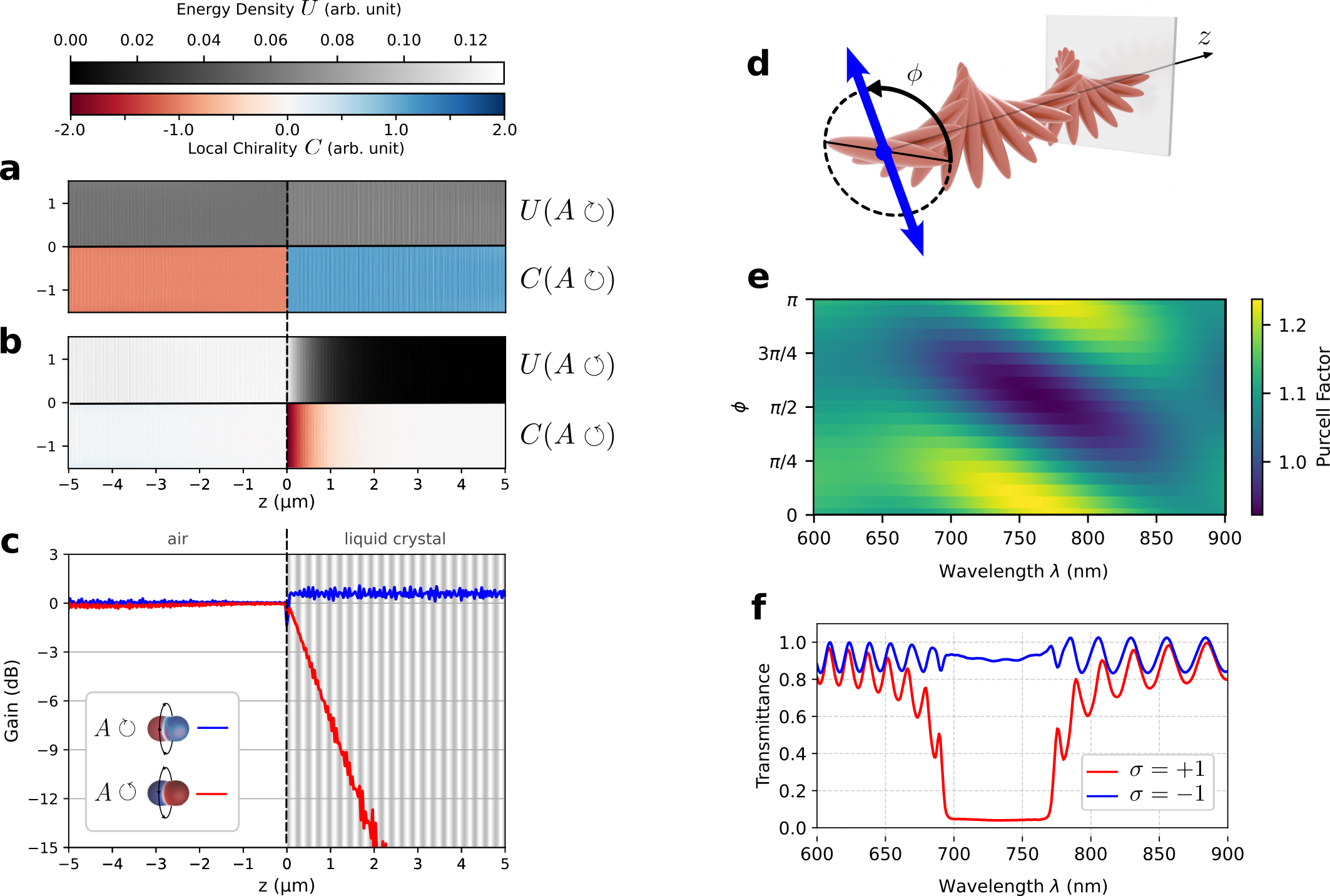} 
    \caption{\textbf{Valley-dependent optical properties at chiral nematic interface.}
(\textbf{a},\textbf{b}) FDTD simulations of the electromagnetic response of a left-handed CLC  ($\chi = -1$) excited by valley excitons distributed at the interface ($z = 0$). $A\circlearrowright$ and $A\circlearrowleft$ excitons cases are shown in (\textbf{a}) and (\textbf{b}), respectively. Upper panels: electromagnetic energy density $U$; lower panels: local optical chirality $C$.
(\textbf{c}) Energy density gain $G = 10 \log ( U / U_0 )$, with $U_0$ the energy density at the interface $z=0^-$.
(\textbf{d}) Schematic of a left-handed CLC with a point-like dipole source at the interface; $\phi$ denotes the angle between the surface local director and the dipole orientation.
(\textbf{e}) Calculated local density of optical states (LDOS) as a function of wavelength and dipole angle $\phi$.
(\textbf{f}) Transmittance spectra for normally incident circularly polarized plane waves with right- ($\sigma = -1$) and left-handed ($\sigma = +1$) polarizations.}
    \label{fig:Fig.4}
\end{figure*}

To estimate the value of the unidirectional coupling rate $\beta$, we propose to probe the penetration depth of the evanescent optical field in case of circular Bragg reflection by FDTD numerical simulations (see methods for details). To do so electromagnetic energy density $U$ and local optical chirality $C$ has been mapped from the electric $\mathbf{E}$ and magnetic $\mathbf{H}$ complex vector fields:

\begin{align}
U=\frac{1}{2}\left(\varepsilon \, |\mathbf{E}|^2+\mu \, |\mathbf{H}|^2\right)
\\[4pt]
C = - \frac{1}{2}\varepsilon_0 \mu_0 \omega \Im \left[\mathbf{E}^* \cdot \mathbf{H}\right]
\end{align}
where $\varepsilon$ and $\mu$ denote the medium-dependent permittivity and permeability, respectively, while $\varepsilon_0$ and $\mu_0$ correspond to their respective values in the vacuum.

Figure \ref{fig:Fig.4}\textbf{a} and \textbf{b} depicts the spatial distribution of $U$ and $C $ generated by circular electric dipoles continuously distributed at the interface ($z = 0$). As expected, when $A\circlearrowright$ excitons are positioned at the surface of a $\chi=1$ CLC, the emitted radiation is essentially unaffected by the presence of the medium and the optical field propagates symmetrically along the $+z$  and $-z$  directions without noticeable attenuation. The calculated optical chirality is -1 (respectively, 1) in the $-z$ (respectively, $+z$) half-space, signature of rotating dipoles emitting opposite circularly polarized light depending on the emission direction as illustrated in Figure \ref{fig:Fig.1}\textbf{e}.  In contrast, for the configuration involving $A\circlearrowleft$ excitons, light emitted toward the $+z$ direction is not allowed to propagate inside the substrate. As a consequence, the energy density $U$ decays rapidly, indicating rapid attenuation of the optical field inside the material. This non-propagating component is then reflected at the interface, undergoes a spin flip, and superimposes with the radiation emitted towards the $-z$  direction. The resulting net optical chirality calculated for the negative $z$ region is therefore equal to zero, as shown in Figure \ref{fig:Fig.4}\textbf{b} as an equal amount of $\sigma$ = 1 and $\sigma$ = -1 light is propagating through this space. The local optical chirality calculated inside the LC displays a pronounced enhancement localized at the interface, reaching values of $C=-2$, as incident and reflected light presenting the same chirality ($C=-1$) occupy the same space, before rapidly decaying to zero over a distance comparable to the optical wavelength.

Figure \ref{fig:Fig.4}\textbf{c} displays the optical gain in dB, defined as $-10\log_{10}(U(z)/U(z=0))$ , as a function of position $z$ for the two configurations discussed above.  One can clearly see that in the case of $A\circlearrowleft$ the electromagnetic field is decreasing exponentially inside the LC. The 3dB loss threshold is reached at a distance $l\approx 500$ nm from the interface. This charasteristic length can be used to estimate the time delay between the photon emission and its possible interaction with the TMD layer after reflection and spin flip by the substrate as $(2l\bar{n})/c \approx 5$ fs. This timescale is two orders of magnitude shorter than the typical excitonic decoherence time reported in WSe$_2$, placing the system comfortably within the regime $\beta > \gamma_\phi$. Furthermore, this result confirms that the Bragg reflection builds up over a propagation length of only 2--3~$\mu$m, ensuring that the effect is robust against variations in the total CLC layer thickness, as long as it remains thicker than a few micrometers.

This confirm that the inter-valley coherence experimentally observed can be mediated by the spin flip reflection inside the LC PBG, allowing the coexisting $A\circlearrowleft$ and $A\circlearrowright$ to coherently combine to form an effective linear dipole emitting linearly polarized light in the far field. 
Nevertheless, while the condition $\beta > \gamma_\phi$ ensures that inter‑valley coherence can be established, it is not sufficient by itself to produce a net linear polarization component in the steady‑state regime. Indeed, although the coherent superposition of $A\circlearrowleft$ and $A\circlearrowright$ excitons gives rise to an effective linear dipole, the orientation of this dipole remains degenerate within the transverse plane, reflecting the intrinsically incoherent nature of the PL originating from the K valley. An additional mechanism is therefore required to lift this orientational degeneracy 

To uncover this missing mechanism, the local density of optical states (LDOS) has been computed for a linear dipole positioned at the CLC interface. As illustrated in Fig.\ref{fig:Fig.4}\textbf{d}, the LDOS is evaluated for all dipole orientations $\phi$ relative to the local director $\textbf{n}$  at the interface. Figure \ref{fig:Fig.4}\textbf{e} shows the resulting Purcell factor as a function of wavelength $\lambda$ and dipole angle $\phi$. At the A exciton energy,  the Purcell enhancement exhibits a clear angular dependence, with an increase of approximately 20$\%$ for $\phi \approx 0, \pi$. This angular selectivity breaks the in‑plane ($x$,$y$) symmetry leading to a preferential emission along this direction. Such anisotropic radiative coupling therefore emerges as a natural candidate for lifting the orientational degeneracy of the linear dipole and mitigating phase randomization in the far field. As the local director at the surface of our CLC substrate is uniform over large distances (compared to 1L-TMD flake size), this symmetry breaking explain why the polarization axis can be controlled by simply rotating the sample.

\section{Conclusion}

In summary, we have shown that chiral nematic liquid crystals provide a simple but efficient photonic environment to mediate intervalley coupling in monolayer TMD at room temperature. By exploiting the intrinsic spin angular momentum flip imposed by chiral Bragg reflection, this platform enables near‑field coupling between excitons residing in opposite K and K' valleys, without the need for nanostructuring or complex cavity architectures.
Polarization‑resolved photoluminescence measurements reveal the onset of intervalley coherence at room temperature, consistent with a rapid chiral‑mediated coupling between excitons in the K and K' valleys. Photons emitted from one valley and selectively reflected by the chiral PBG experience a spin angular momentum flip, which can in turn promote coherent radiative coupling to excitons in the opposite valley while preserving polarization and phase. Because this chiral‑mediated coupling occurs on timescales faster than room‑temperature dephasing, a coherent superposition of valley excitons can be sustained, resulting in the emergence of a finite linear polarization in the far field.
More broadly, our results establish CLC as a versatile and scalable platform for engineering coherent valley‑dependent light–matter interactions. By relying on intrinsic material chirality rather than finely tuned nanofabrication, this approach offers a robust route toward room‑temperature valleytronic functionalities based on controlled coherent emission from valleys with opposite Berry curvature, and opens new perspectives for chiral and topological photonics in two‑dimensional semiconductors. Although the present demonstration employs CLCs in their glassy solid state, the underlying physics is fully generalizable to their liquid crystalline phase, in which the PBG spectral position can be dynamically tuned through external stimuli such as electric fields, temperature, or optical excitation. This intrinsic tunability, well established in display and photonic technologies, would provide an additional degree of freedom for the active control of valley-selective emission, opening a route toward electrically switchable valleytronic devices operating at room temperature.

\section{Methods}

\subsection*{Sample fabrication}
The CLC used in this study consist of a siloxane cyclic backbone functionalized with two types of side chains : an achiral and a chiral mesogen (Wacker Chemie GmbH). The cholesteric phase forms between 180–210°C (clearing‑temperature range) and 40–50°C (glass‑transition range).  For this work, we prepared 2 types of mixtures to position the center of the PBG at 735 nm, matching the TMD WSe$_2$ PL (sample \textsc{in})  and at 550 nm, far from the PL range (sample \textsc{out}).
The CLC layers were formed by sandwiching the material between two glass substrates separated by Mylar 15$\mu$m-thick spacers. No alignment coatings or surface‑conditioning treatments were applied since the preferential anchoring of these molecules with glass is planar. The assembled cells were annealed at 140°C for 2 minutes to ensure planar Grandjean organization, after which they were abruptly quenched at room-temperature. This rapid thermal drop immobilizes the CLC structure, converting the viscous material into a glassy solid. To enable clean detachment of the top glass substrate and to produce a smooth exposed interface, the quenched samples were subsequently cooled to approximately 0°C before opening. 
Monolayer WSe$ _2 $ flakes were obtained by mechanical exfoliation from a commercial bulk crystal (from \textit{2D Semiconductors}) and directly transferred using a polydimethylsiloxane (PDMS) polymer stamp~\cite{castellanos2014deterministic} onto the CLC film.

\subsection{Optical measurements}
All the optical measurements were conducted in ambient environment at room temperature.
They were performed using an inverted confocal microscope (\textit{Nikon Eclipse Ti2}) equipped with a 100×/0.9 objective (\textit{Olympus MPlanFL N}). The excitation beam was provided by a supercontinuum laser source (\textit{YSL Photonics SC-Pro7}), spectrally filtered using a tunable bandpass filter (\textit{Poly-RED-A5 }from\textit{ Spectrolight Inc}) to isolate a wavelength of 638nm with a bandwidth of 3nm. The excitation was pulsed at a repetition rate of 50MHz, with a pulse duration of 1ns.
PL was collected by placing a 300$\mu$m‑diameter multimode optical fiber at the conjugate image plane, with the fiber core acting as the confocal pinhole. A non‑polarization‑maintaining fiber was used to avoid any polarization‑dependent response of the detection system. The excitation light was filtered with a longpass filter (\textit{FELH0650 }from\textit{ Thorlabs}).The PL signal was analyzed using a spectrometer (\textit{Andor Shamrock}) equipped with a camera (\textit{Andor IDus}), cooled to -60$^\circ$C. Polarization-resolved measurements was realized with a zero-order achromatic quarter waveplate (\textit{AQWP05M-580 }from\textit{ Thorlabs}) and a linear polarizer with a 600-1100nm operating range  (\textit{LPNIRE100-B }from\textit{ Thorlabs}).

\subsection{Data analysis}
\subsubsection{Lorentzian fit of PL}

To remove from the PL spectra the contribution of intrinsically unpolarized low-energy excitons, the high-energy part of each measured spectrum is fitted with a Lorentzian function corresponding to the bright exciton resonance. The fitted bright-exciton peak is then subtracted from the total photoluminescence signal to isolate the lower-energy excitonic contribution, as shown in Figure \ref{fig:Fig.1}\textbf{d}. This residual low-energy contribution reflecting the presence of multiple low-energy excitonic resonances at different energies and is not taken into account in this study.\cite{shradha20262d,lamsaadi2025exciton} In Figure \ref{fig:Fig.3}, the PL intensities are extracted from the integrated area under the fitted Lorentzian curve related to the bright A-exciton resonance only. Complete sets of recorded PL with the corresponding fits are presented in Figure SI1 of the Supplementary Information.

\subsubsection{Calibration-free measurements of circular polarization populations}

The polarization state of the emitted PL is intrinsically a combination of left‑ and right‑handed circularly polarized components. When these two components are partially coherent, a linear polarization characterized by a degree $P$  and an orientation angle $\alpha$  must also be taken into account. The Stokes vector of the PL emission can then be written as :
\begin{equation}
\mathbf{S}_{\textsc{pl}}
=
\begin{pmatrix}
1 \\
P \cos 2\alpha \\
P \sin 2\alpha \\
DoCP
\end{pmatrix}
\end{equation}
where the degree of circular polarization $DoCP = (I_{\sigma^+} - I_{\sigma^-}) / (I_{\sigma^+} + I_{\sigma^-})$ quantifies the imbalance between the intensities $I_{\sigma^+}$ and  $I_{\sigma^-}$ of the right- and the left-handed circularly polarized components, respectively.

In the detection line presented in Figure \ref{fig:Fig.2}, the photoluminescence successively passes through a $\lambda/4$ plate oriented at $\pm 45^\circ$, a beam splitter (BS), and a linear analyzer (A) oriented at 0$^\circ$  or 90$^\circ$. The measured intensity corresponds to the first component of the Stokes vector obtained after propagation through the detection line : 
\begin{equation}
I = \left[ 
\mathbf M_{\mathrm{A}}^{0^\circ,\,90^\circ} \,
\mathbf{M}_{\mathrm{BS}} \,
\mathbf M_{\lambda/4}^{\pm 45^\circ}\,
\right]_0
\end{equation}
where $\mathbf M_{\mathrm{A}}^{0^\circ,\,90^\circ}$, $\mathbf{M}_{\mathrm{BS}}$ and $\mathbf M_{\lambda/4}^{\pm 45^\circ}$ are de Mueller matrices of corresponding optical elements. 

In this framework, a straightforward calculus detailed in the Supplementary Information leads to these relations :
\begin{align}
I_{\sigma^+} &= I^{+45^\circ}_{0^\circ} + I^{-45^\circ}_{90^\circ}, \\
I_{\sigma^-} &= I^{+45^\circ}_{90^\circ} + I^{-45^\circ}_{0^\circ}
\end{align}

This crossed-sum procedure exactly cancels the differential transmission of the beam splitter, making the retrieval of $I_{\sigma^+}$ and $I_{\sigma^+}$ independent of the polarization dependent transmission coefficients of the BS, and of a possible partial coherence between the right- and left-handed components.

\subsubsection{Correction of beam-splitter--induced distortions in linear polarization analysis}

The analysis of the linear polarization is affected by the residual polarization-dependent response of the beam splitter (BS) and therefore requires an explicit correction.
In the linear polarization measurements, the photoluminescence is analyzed using a rotating linear polarizer placed after the BS (See Fig. \ref{fig:Fig.3}). The BS is modeled within the Mueller formalism as a linear diattenuator characterized by two different transmission coefficients for the $s$- and $p$-polarized components, $t_s$ and $t_p$, respectively.

In the absence of the BS, a rotating linear analyzer would yield the angular dependence of the detected intensity :
\begin{equation}
I_0(\theta)
=
\frac{1}{2}
\left[
1 + P \cos 2(\theta - \alpha)
\right]
\end{equation}
where $\theta$ measures the rotation of the analyzer with respect to the $x$ axis.

In the actual experimental configuration, i.e. in the presence of the BS, the angular dependence of the detected intensity explicitly depends on the BS transmission coefficients $t_s$ and $t_p$, with $\delta = ts-tp$ (see Supplementary Information for details):
\begin{equation}
I_{\mathrm{out}}(\theta)
=
\frac{1}{4}
\Big[
1 + \delta P \cos 2\alpha
+ \big( \delta + P \cos 2\alpha \big)\cos 2\theta
+ 2\sqrt{t_s t_p}\, P \sin 2\alpha \sin 2\theta
\Big]
\end{equation} 
A quantitative reconstruction of the intrinsic linear polarization therefore requires an independent calibration of the BS response. This calibration is performed using the reference sample, emitting intrinsically unpolarized photoluminescence and for which the incident Stokes vector reduces to $\mathbf{S}_{\mathrm{ref}} = (1,0,0,0)$. In this case, the observed angular modulation of the detected intensity originates solely from the BS diattenuation, enabling an unambiguous determination of the coefficients $t_s$ and $t_p$. Once calibrated, these coefficients are treated as fixed parameters, allowing for a unique inversion of the measured angular dependence and a quantitative reconstruction of the intrinsic PL intensity $I_0(\theta)$. The corresponding raw and corrected polar representations are shown in Figure SI2.

\subsection{Numerical simulations}
Simulations were performed with the finite-difference time-domain (FDTD) method, using the \textit{Meep} open-source software package \cite{oskooi2010meep}. The three‑dimensional FDTD simulations shown in Figure \ref{fig:Fig.4} were conducted on a CLC domain of dimensions $3 \times 3 \times 8\,\mu\text{m}^3$, employing a uniform spatial mesh with a resolution of 10nm. To suppress undesired boundary reflections, perfectly matched layers (PML) with an additional thickness of $1\,\mu\text{m}$ were implemented along all the three spatial directions.

\subsection*{Disclosures}
The authors declare that there are no financial interests, commercial affiliations, or other potential conflicts of interest that could have influenced the objectivity of this research or the writing of this paper

\subsection*{Data availability}
All data supporting the findings of this study are included within the Article and its Supplementary Information, with additional data available from the corresponding author upon reasonable request. 
Correspondence and requests for materials should be addressed to J.M.P. (email: jean-marie.poumirol@cemes.fr) or G.A. (email: gonzague.agez@cemes.fr)

\subsection*{Acknowledgements}
This study has been supported through the grant NanoX n° ANR-17-EURE-0009 in the framework of the \textit{Programme des Investissements d’Avenir}.
The authors acknowledge the PyMoDAQ open-source community for providing tools used in data acquisition and control \cite{weber2021pymodaq}.


\bibliography{biblio}   

@article{zhu2014anomalously,
  title={Anomalously robust valley polarization and valley coherence in bilayer WS2},
  author={Zhu, Bairen and Zeng, Hualing and Dai, Junfeng and Gong, Zhirui and Cui, Xiaodong},
  journal={Proceedings of the National Academy of Sciences},
  volume={111},
  number={32},
  pages={11606--11611},
  year={2014},
  publisher={National Academy of Sciences}
}

@article{jones2013optical,
  title={Optical generation of excitonic valley coherence in monolayer WSe$_2$},
  author={Jones, Aaron M and Yu, Hongyi and Ghimire, Nirmal J and Wu, Sanfeng and Aivazian, Grant and Ross, Jason S and Zhao, Bo and Yan, Jiaqiang and Mandrus, David G and Xiao, Di and others},
  journal={Nature nanotechnology},
  volume={8},
  number={9},
  pages={634--638},
  year={2013},
  publisher={Nature Publishing Group UK London}
}

@article{shradha20262d,
  title={2D excitonics with atomically thin lateral heterostructures},
  author={Shradha, S and Rosati, R and Lamsaadi, H and Picker, J and Paradisanos, I and Hossain, Md T and Krelle, L and Oswald, LF and Engel, N and Markina, DI and others},
  journal={Reports on Progress in Physics},
  volume={89},
  number={4},
  pages={046501},
  year={2026},
  publisher={IOP Publishing}
}

@article{huang2016probing,
  title={Probing the origin of excitonic states in monolayer WSe$_2$},
  author={Huang, Jiani and Hoang, Thang B and Mikkelsen, Maiken H},
  journal={Scientific reports},
  volume={6},
  number={1},
  pages={22414},
  year={2016},
  publisher={Nature Publishing Group UK London}
}

@article{yan2015valley,
  title={Valley depolarization in monolayer WSe$_2$},
  author={Yan, Tengfei and Qiao, Xiaofen and Tan, Pingheng and Zhang, Xinhui},
  journal={Scientific reports},
  volume={5},
  number={1},
  pages={15625},
  year={2015},
  publisher={Nature Publishing Group UK London}
}

@article{hao2016direct,
  title={Direct measurement of exciton valley coherence in monolayer WSe$_2$},
  author={Hao, Kai and Moody, Galan and Wu, Fengcheng and Dass, Chandriker Kavir and Xu, Lixiang and Chen, Chang-Hsiao and Sun, Liuyang and Li, Ming-Yang and Li, Lain-Jong and MacDonald, Allan H and others},
  journal={Nature Physics},
  volume={12},
  number={7},
  pages={677--682},
  year={2016},
  publisher={Nature Publishing Group UK London}
}

@article{cadiz2022imaging,
  title={Imaging the effect of high photoexcited densities on valley polarization and coherence in MoS$_2$ monolayers},
  author={Cadiz, F and Gerl, S and Taniguchi, T and Watanabe, K},
  journal={npj 2D Materials and Applications},
  volume={6},
  number={1},
  pages={27},
  year={2022},
  publisher={Nature Publishing Group UK London}
}

@article{gupta2023observation,
  title={Observation of $\approx$  100$\%$ valley-coherent excitons in monolayer MoS$_2$ through giant enhancement of valley coherence time},
  author={Gupta, Garima and Watanabe, Kenji and Taniguchi, Takashi and Majumdar, Kausik},
  journal={Light: Science \& Applications},
  volume={12},
  number={1},
  pages={173},
  year={2023},
  publisher={Nature Publishing Group UK London}
}

@article{lorchat2018room,
  title={Room-temperature valley polarization and coherence in transition metal dichalcogenide--graphene van der Waals heterostructures},
  author={Lorchat, Etienne and Azzini, Stefano and Chervy, Thibault and Taniguchi, Takashi and Watanabe, Kenji and Ebbesen, Thomas W and Genet, Cyriaque and Berciaud, St{\'e}phane},
  journal={ACS photonics},
  volume={5},
  number={12},
  pages={5047--5054},
  year={2018},
  publisher={ACS Publications}
}

@article{lamsaadi2025exciton,
  title={Exciton Collimation, Focusing and Trapping Using Complex Transition Metal Dichalcogenide Lateral Heterojunctions},
  author={Lamsaadi, Hassan and Cuche, Aurelien and Agez, Gonzague and Paradisanos, Ioannis and Beret, Dorian and Lombez, Laurent and Renucci, Pierre and Lagarde, Delphine and Marie, Xavier and Gan, Ziyang and others},
  journal={Advanced Optical Materials},
  volume={13},
  number={10},
  pages={2403009},
  year={2025},
  publisher={Wiley Online Library}
}

@article{cadiz2018exciton,
  title={Exciton diffusion in WSe2 monolayers embedded in a van der Waals heterostructure},
  author={Cadiz, Fabian and Robert, C{\'e}dric and Courtade, Emmanuel and Manca, Marco and Martinelli, L and Taniguchi, T and Watanabe, K and Amand, Thierry and Rowe, ACH and Paget, D and others},
  journal={Applied Physics Letters},
  volume={112},
  number={15},
  year={2018},
  publisher={AIP Publishing}
}

@article{yan2014photoluminescence,
  title={Photoluminescence properties and exciton dynamics in monolayer WSe$_2$},
  author={Yan, Tengfei and Qiao, Xiaofen and Liu, Xiaona and Tan, Pingheng and Zhang, Xinhui},
  journal={Applied Physics Letters},
  volume={105},
  number={10},
  year={2014},
  publisher={AIP Publishing}
}

@article{godde2016exciton,
  title={Exciton and trion dynamics in atomically thin MoSe$_2$ and WSe$_2$: Effect of localization},
  author={Godde, T and Schmidt, D and Schmutzler, J and A{\ss}mann, M and Debus, J and Withers, F and Alexeev, EM and Del Pozo-Zamudio, O and Skrypka, OV and Novoselov, KS and others},
  journal={Physical Review B},
  volume={94},
  number={16},
  pages={165301},
  year={2016},
  publisher={APS}
}

@article{brem2020phonon,
  title={Phonon-assisted photoluminescence from indirect excitons in monolayers of transition-metal dichalcogenides},
  author={Brem, Samuel and Ekman, August and Christiansen, Dominik and Katsch, Florian and Selig, Malte and Robert, Cedric and Marie, Xavier and Urbaszek, Bernhard and Knorr, Andreas and Malic, Ermin},
  journal={Nano letters},
  volume={20},
  number={4},
  pages={2849--2856},
  year={2020},
  publisher={ACS Publications}
}

@article{li2018revealing,
  title={Revealing the biexciton and trion-exciton complexes in BN encapsulated WSe$_2$},
  author={Li, Zhipeng and Wang, Tianmeng and Lu, Zhengguang and Jin, Chenhao and Chen, Yanwen and Meng, Yuze and Lian, Zhen and Taniguchi, Takashi and Watanabe, Kenji and Zhang, Shengbai and others},
  journal={Nature communications},
  volume={9},
  number={1},
  pages={3719},
  year={2018},
  publisher={Nature Publishing Group UK London}
}

@article{zhou2017probing,
  title={Probing dark excitons in atomically thin semiconductors via near-field coupling to surface plasmon polaritons},
  author={Zhou, You and Scuri, Giovanni and Wild, Dominik S and High, Alexander A and Dibos, Alan and Jauregui, Luis A and Shu, Chi and De Greve, Kristiaan and Pistunova, Kateryna and Joe, Andrew Y and others},
  journal={Nature nanotechnology},
  volume={12},
  number={9},
  pages={856--860},
  year={2017},
  publisher={Nature Publishing Group UK London}
}

@article{castellanos2014deterministic,
  title={Deterministic transfer of two-dimensional materials by all-dry viscoelastic stamping},
  author={Castellanos-Gomez, Andres and Buscema, Michele and Molenaar, Rianda and Singh, Vibhor and Janssen, Laurens and Van Der Zant, Herre SJ and Steele, Gary A},
  journal={2D Materials},
  volume={1},
  number={1},
  pages={011002},
  year={2014},
  publisher={IOP Publishing}
}

@article{rivera2018interlayer,
  title={Interlayer valley excitons in heterobilayers of transition metal dichalcogenides},
  author={Rivera, Pasqual and Yu, Hongyi and Seyler, Kyle L and Wilson, Nathan P and Yao, Wang and Xu, Xiaodong},
  journal={Nature nanotechnology},
  volume={13},
  number={11},
  pages={1004--1015},
  year={2018},
  publisher={Nature Publishing Group UK London}
}

@article{schaibley2016valleytronics,
  title={Valleytronics in 2D materials},
  author={Schaibley, John R and Yu, Hongyi and Clark, Genevieve and Rivera, Pasqual and Ross, Jason S and Seyler, Kyle L and Yao, Wang and Xu, Xiaodong},
  journal={Nature Reviews Materials},
  volume={1},
  number={11},
  pages={1--15},
  year={2016},
  publisher={Nature Publishing Group}
}

@article{unuchek2018room,
  title={Room-temperature electrical control of exciton flux in a van der Waals heterostructure},
  author={Unuchek, Dmitrii and Ciarrocchi, Alberto and Avsar, Ahmet and Watanabe, Kenji and Taniguchi, Takashi and Kis, Andras},
  journal={Nature},
  volume={560},
  number={7718},
  pages={340--344},
  year={2018},
  publisher={Nature Publishing Group UK London}
}

@article{smolenski_tuning_2016,
  title={Tuning valley polarization in a WSe$_2$ monolayer with a tiny magnetic field},
  author={Smole{\'n}ski, T and Goryca, M and Koperski, M and Faugeras, C and Kazimierczuk, T and Bogucki, A and Nogajewski, K and Kossacki, P and Potemski, M},
  journal={Physical Review X},
  volume={6},
  number={2},
  pages={021024},
  year={2016},
  publisher={APS}
}

@article{mak2016photonics,
  title={Photonics and optoelectronics of 2D semiconductor transition metal dichalcogenides},
  author={Mak, Kin Fai and Shan, Jie},
  journal={Nature Photonics},
  volume={10},
  number={4},
  pages={216--226},
  year={2016},
  publisher={Nature Publishing Group UK London}
}

@article{xu2014spin,
  title={Spin and pseudospins in layered transition metal dichalcogenides},
  author={Xu, Xiaodong and Yao, Wang and Xiao, Di and Heinz, Tony F},
  journal={Nature Physics},
  volume={10},
  number={5},
  pages={343--350},
  year={2014},
  publisher={Nature Publishing Group UK London}
}

@article{mak2012control,
  title={Control of valley polarization in monolayer MoS$_2$ by optical helicity},
  author={Mak, Kin Fai and He, Keliang and Shan, Jie and Heinz, Tony F},
  journal={Nature nanotechnology},
  volume={7},
  number={8},
  pages={494--498},
  year={2012},
  publisher={Nature Publishing Group UK London}
}

@article{zeng2012valley,
  title={Valley polarization in MoS$_2$ monolayers by optical pumping},
  author={Zeng, Hualing and Dai, Junfeng and Yao, Wang and Xiao, Di and Cui, Xiaodong},
  journal={Nature nanotechnology},
  volume={7},
  number={8},
  pages={490--493},
  year={2012},
  publisher={Nature Publishing Group UK London}
}

@article{cao2012valley,
  title={Valley-selective circular dichroism of monolayer molybdenum disulphide},
  author={Cao, Ting and Wang, Gang and Han, Wenpeng and Ye, Huiqi and Zhu, Chuanrui and Shi, Junren and Niu, Qian and Tan, Pingheng and Wang, Enge and Liu, Baoli and others},
  journal={Nature communications},
  volume={3},
  number={1},
  pages={887},
  year={2012},
  publisher={Nature Publishing Group UK London}
}

@article{rong_spin-valley_2023,
  title={Spin-valley Rashba monolayer laser},
  author={Rong, Kexiu and Duan, Xiaoyang and Wang, Bo and Reichenberg, Dror and Cohen, Assael and Liu, Chieh-li and Mohapatra, Pranab K and Patsha, Avinash and Gorovoy, Vladi and Mukherjee, Subhrajit and others},
  journal={Nature Materials},
  volume={22},
  number={9},
  pages={1085--1093},
  year={2023},
  publisher={Nature Publishing Group UK London}
}

@article{rong_photonic_2020,
  title={Photonic Rashba effect from quantum emitters mediated by a Berry-phase defective photonic crystal},
  author={Rong, Kexiu and Wang, Bo and Reuven, Avi and Maguid, Elhanan and Cohn, Bar and Kleiner, Vladimir and Katznelson, Shaul and Koren, Elad and Hasman, Erez},
  journal={Nature Nanotechnology},
  volume={15},
  number={11},
  pages={927--933},
  year={2020},
  publisher={Nature Publishing Group UK London}
}

@article{loren_circular_2023,
  title={Circular dichroism induction in WS$_2$ by a chiral plasmonic metasurface},
  author={Lor{\'e}n, Fernando and Genet, Cyriaque and Martin-Moreno, Luis},
  journal={Optical Materials Express},
  volume={13},
  number={11},
  pages={3366--3375},
  year={2023},
  publisher={Optica Publishing Group}
}

@article{guddala_valley_2019,
  title={Valley selective optical control of excitons in 2D semiconductors using a chiral metasurface},
  author={Guddala, S and Bushati, R and Li, M and Khanikaev, AB and Menon, VM},
  journal={Optical Materials Express},
  volume={9},
  number={2},
  pages={536--543},
  year={2019},
  publisher={Optical Society of America}
}

@article{jimenez-galan_lightwave_2020,
  title={Lightwave control of topological properties in 2D materials for sub-cycle and non-resonant valley manipulation},
  author={Jim{\'e}nez-Gal{\'a}n, Alvaro and Silva, REF and Smirnova, O and Ivanov, M},
  journal={Nature Photonics},
  volume={14},
  number={12},
  pages={728--732},
  year={2020},
  publisher={Nature Publishing Group UK London}
}

@article{rafayelyan2017ultrabroadband,
  title={Ultrabroadband gradient-pitch Bragg-Berry mirrors},
  author={Rafayelyan, Mushegh and Agez, Gonzague and Brasselet, Etienne},
  journal={Physical Review A: Atomic, molecular, and optical physics [1990-2015]},
  volume={96},
  number={4},
  pages={043862},
  year={2017}
}

@article{barboza2016berry,
  title={Berry phase of light under Bragg reflection by chiral liquid-crystal media},
  author={Barboza, Raouf and Bortolozzo, Umberto and Clerc, Marcel G and Residori, Stefania},
  journal={Physical Review Letters},
  volume={117},
  number={5},
  pages={053903},
  year={2016},
  publisher={APS}
}

@article{wang2019fully,
  title={Fully chiral light emission from CsPbX3 perovskite nanocrystals enabled by cholesteric superstructure stacks},
  author={Wang, Chun-Ta and Chen, Keqiang and Xu, Ping and Yeung, Fion and Kwok, Hoi-Sing and Li, Guijun},
  journal={Advanced Functional Materials},
  volume={29},
  number={35},
  pages={1903155},
  year={2019},
  publisher={Wiley Online Library}
}

@article{wang2025direct,
  title={Direct Identification of Valley Coherence and Its Manipulation in Monolayer Two-Dimensional Semiconductor},
  author={Wang, Haonan and Shinokita, Keisuke and Watanabe, Kenji and Taniguchi, Takashi and Konabe, Satoru and Matsuda, Kazunari},
  journal={ACS nano},
  volume={19},
  number={23},
  pages={21484--21491},
  year={2025},
  publisher={ACS Publications}
}

@article{oskooi2010meep,
  title={MEEP: A flexible free-software package for electromagnetic simulations by the FDTD method},
  author={Oskooi, Ardavan F and Roundy, David and Ibanescu, Mihai and Bermel, Peter and Joannopoulos, John D and Johnson, Steven G},
  journal={Computer Physics Communications},
  volume={181},
  number={3},
  pages={687--702},
  year={2010},
  publisher={Elsevier}
}

@article{weber2021pymodaq,
  title={PyMoDAQ: An open-source Python-based software for modular data acquisition},
  author={Weber, SJ},
  journal={Review of Scientific Instruments},
  volume={92},
  number={4},
  year={2021},
  publisher={AIP Publishing}
}
\bibliographystyle{spiejour}   


\vspace{2ex}\noindent\textbf{Hassan Lamsaadi} is a postdoctoral researcher at the University of Toulouse. He received his M.Sc. degree in Physics from the University of Toulouse in 2022. He earned his Ph.D. in Nano-Optics from the University of Toulouse in 2025. His current research interests include nanophotonics, nano-optics, and quantum technologies, with a particular focus on light–matter interactions in low-dimensional nanostructures and emerging quantum photonic systems.

\vspace{2ex}\noindent\textbf{Jean Marie Poumirol} is a researcher at CNRS/University of Toulouse and team leader of the NeO group at CEMES laboratory. He received his PhD degree in nanoscience from the Institut National des Sciences Appliquées de Toulouse in 2011. His current research interests focus on excitonic and plasmonic properties in low dimensional materials.

\vspace{2ex}\noindent\textbf{Gonzague Agez} is an associate professor at the University of Toulouse. He received his PhD degree in Laser, Atoms $\&$ Molecules from the Université de Lille in 2005. His current research interests include nonlinear optics, liquid crystals, and nano-optics.

\vspace{1ex}
\noindent Biographies of the other authors are not available.

\listoffigures

\end{spacing}

\end{document}